\documentclass[runningheads]{llncs}
\usepackage[T1]{fontenc}
\usepackage{multirow}
\usepackage{amsmath}
\usepackage{hyperref}
\usepackage{cite}
\usepackage{graphicx,verbatim}
\usepackage{color}
\usepackage{marvosym}
\usepackage[marginal]{footmisc}

\begin{document}
%
% \title{ Cross-Scanner Cervical Cancer Screening via Style-Conditioned Feature Augmentation }
\title{ LSA: Latent Style Augmentation Towards Stain-Agnostic Cervical Cancer Screening}
%
\begin{comment}  %% Removed for anonymized MICCAI 2025 submission

\end{comment}

% \author{Anonymized Authors}  %% Added for anonymized MICCAI 2025 submission
% \authorrunning{Anonymized Author et al.}
% \institute{Anonymized Affiliations \\
%     \email{email@anonymized.com}}

\author{Jiangdong Cai\inst{1}* \and Haotian Jiang\inst{1}* \and Zhenrong Shen\inst{3} \and Yonghao Li\inst{1} \and Honglin Xiong\inst{1} \and Lichi Zhang\inst{3} \and Qian Wang\inst{1,2}\textsuperscript{(\Letter)} }
%index{Cai,Jiangdong; Xiong,Honglin; Cao,Maosong; Liu,Luyan; Zhang,Lichi; Wang,Qian.}

% \institute{School of Biomedical Engineering \and State Key Laboratory of Advanced Medical Materials and Devices, ShanghaiTech University, Shanghai, China} \and {Shanghai Clinical Research and Trial Center, Shanghai, China} \and {School of Biomedical Engineering, Shanghai Jiao Tong University, Shanghai, China}

\institute{{School of Biomedical Engineering \& State Key Laboratory of Advanced Medical Materials and Devices, ShanghaiTech University, Shanghai, China} \and {Shanghai Clinical Research and Trial Center, Shanghai, China} \and {School of Biomedical Engineering, Shanghai Jiao Tong University, Shanghai, China}
}
\authorrunning{J. Cai et al.}
\maketitle              % typeset the header of the contribution

\begin{abstract}

The deployment of computer-aided diagnosis systems for cervical cancer screening using whole slide images (WSIs) faces critical challenges due to domain shifts caused by staining variations across different scanners and imaging environments.
While existing stain augmentation methods improve patch-level robustness, they fail to scale to WSIs due to two key limitations: (1) inconsistent stain patterns when extending patch operations to gigapixel slides, and (2) prohibitive computational/storage costs from offline processing of augmented WSIs.
To address this, we propose \textbf{L}atent \textbf{S}tyle \textbf{A}ugmentation (\textbf{LSA}), a framework that performs efficient, online stain augmentation directly on WSI-level latent features.
We first introduce \textbf{WSAug}, a \textbf{W}SI-level \textbf{s}tain \textbf{aug}mentation method ensuring consistent stain across patches within a WSI. 
Using offline-augmented WSIs by WSAug, we design and train Stain Transformer, which can simulate targeted style in the latent space, efficiently enhancing the robustness of the WSI-level classifier.
We validate our method on a multi-scanner WSI dataset for cervical cancer diagnosis.
Despite being trained on data from a single scanner, our approach achieves significant performance improvements on out-of-distribution data from other scanners. Code will be available at \url{https://github.com/caijd2000/LSA}.

\keywords{Cervical Cancer Diagnosis \and Whole Slide Image  \and Domain Shift \and Data Augmentation.}
% Authors must provide keywords and are not allowed to remove this Keyword section.

\end{abstract}

\footnote{* These authors contributed equally.}

\section{Introduction}

Cervical cancer is a significant global health concern, affecting millions of women each year \cite{gultekin2020world}.
With the progress of digital whole slide image (WSI) scanning instruments and AI technologies, numerous attempts have been made to develop computer-aided diagnosis systems for cervical cancer screening.
Multi-stage frameworks \cite{cheng2021robust,wang2024artificial} based on patch-level detection or classification are extensively utilized, where instance-level models \cite{jiang2023donet,cai2023progressive} are trained with patch-level annotations to localize abnormalities before aggregating predictions at the WSI level. 
Yet, their effectiveness heavily relies on large-scale, high-quality annotations (e.g., lesion boundaries of abnormal cells), which are labor-intensive to acquire and rarely available for cervical cancer datasets. 
To circumvent fine-grained supervision, recent studies adopt pre-trained encoders with multiple instance learning (MIL) \cite{shao2021transmil,yang2024mambamil,ilse2018attention}, leveraging features from pathology foundation models \cite{hoptimus0,huang2023visual,chen2024towards} to infer WSI-level diagnoses through feature relationship modeling \cite{li2024large,cao2023detection}.
% \textcolor{red}{Quite sloppy to handle these citations. Don't need to put too many citatiosn together for a single place.}

Despite the rapid development of AI-based systems, the diagnostic performance on WSIs is often compromised by domain shifts, especially discrepancies in staining variations across different scanners and physical environments \cite{jahanifar2023domaingeneralizationcomputationalpathology}.
% Addressing this challenge is critical to ensuring reliable and equitable diagnosis across diverse healthcare settings.
Existing methods can be broadly categorized into stain normalization \cite{salehi2020pix2pix,bentaieb2017adversarial,shaban2019staingan} and stain augmentation  \cite{wagner2021structure,shen2022randstainna,zhang2023learning}, which achieve notable success in patch-level model training by aligning color distributions or simulating staining variations.

% While stain augmentation has proven effective for scanner-invariant patch-level feature learning, its extension to WSIs faces three critical barriers. First, the gigapixel resolution of WSIs leads to the absence of fine-grained annotations in most WSI datasets, which restricts stain augmentation to supervised learning on sparsely annotated patches \cite{madusanka2023impact}.
% Second, while unsupervised pre-training mitigates patch-level domain shifts, the absence of online WSI augmentation perpetuates source bias in the downstream fine-tuning of slide-level classifiers, as domain discrepancies persist during feature aggregation. 
% % \textcolor{red}{I don't fully understand your argument here, especially considering many large models claim they are so good.}
% Third, brute-force extension of patch-level methods to WSIs incurs prohibitive storage costs (e.g., generating multiple augmented slide copies) and disrupts slide-wide staining coherence—a prerequisite for diagnosing spatially distributed abnormalities.

While stain augmentation has demonstrated effectiveness for scanner-invariant patch-level feature learning, its extension to WSIs encounters two critical challenges.
First, directly applying patch-level methods to WSIs disrupts the staining coherence across the entire WSI, affecting the training of WSI classifiers.
Second, the gigapixel resolution of WSIs introduces prohibitive computational and storage costs when generating multiple augmented slide copies offline, posing a significant barrier to practical implementation.

To overcome this challenge, we propose a novel Latent Style Augmentation (LSA) framework that performs data augmentation directly on WSI-level features. 
LSA enables the WSI-level classifier to maintain robustness across different stain styles, without explicitly generating WSIs or repeatedly extracting features.
% SCLA can be trained on the features extracted by various pathology foundation models without fine-tuning. 
Models trained with LSA demonstrate excellent cross-scanner performance on both in-distribution (ID) and out-of-distribution (OOD) data.
The main contributions of our work can be summarized as follows:
\begin{itemize}
    \item We propose the WSI-level stain augmentation method WSAug to ensure staining consistency across patches. 
    \item The proposed LSA framework transforms WSI-level features into arbitrary staining styles, enabling WSI-level classifiers to remain robust across diverse stain styles. 
    \item Experiments on a multi-scanner cervical cancer dataset demonstrate that LSA allows training on a single scanner’s data while maintaining consistent diagnostic performance across other scanners.
\end{itemize}

\section{Method}

\begin{figure*}[t]
    \centering
    \includegraphics[width=\textwidth]{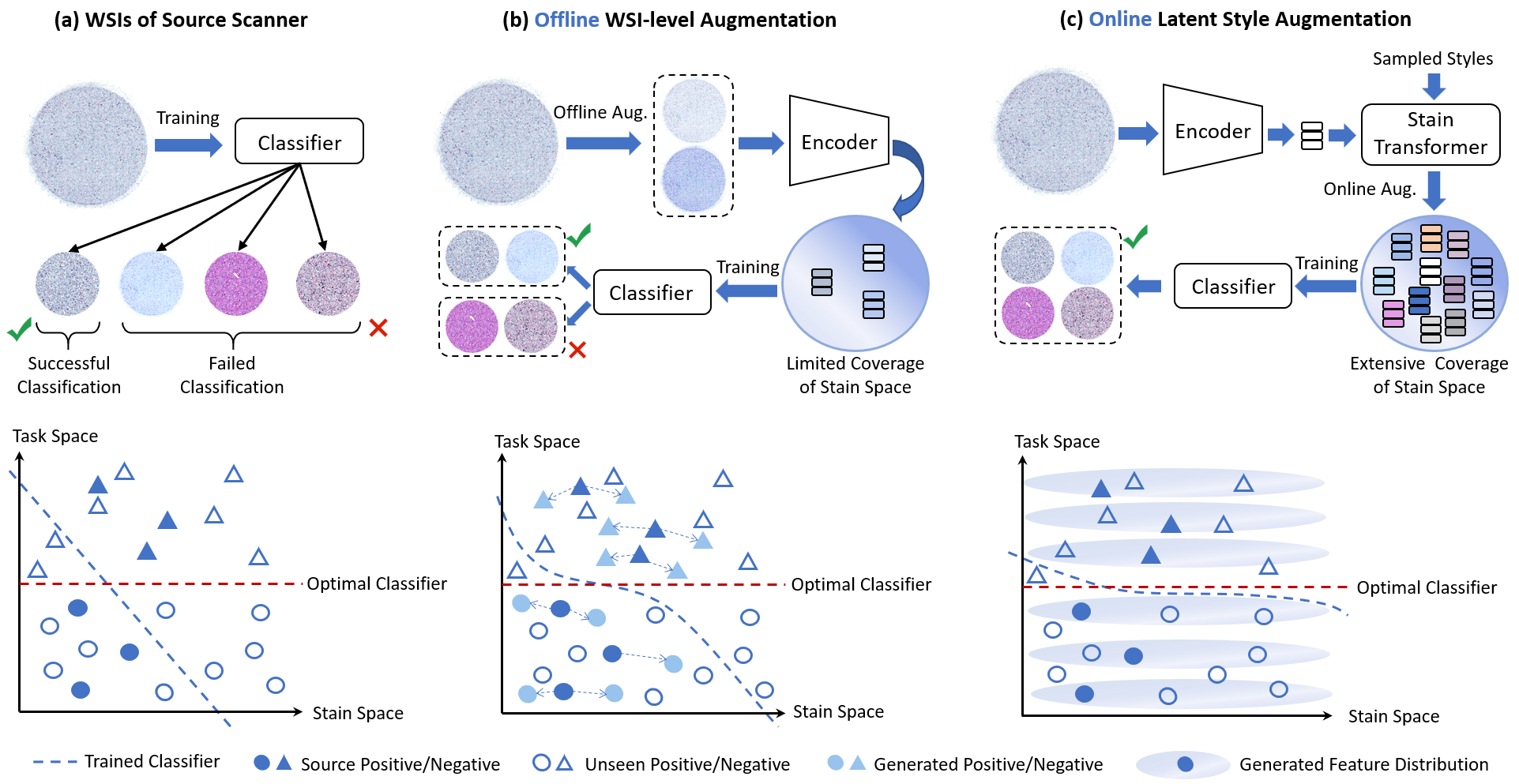}
    \caption{Overview of method. (a) Training with only WSIs of the source scanner limits the classifier’s ability to handle unseen stain types. (b) Offline WSI-level data augmentation broaders the capability but cannot cover the entire spectrum of stain styles. (c) Online latent style augmentation can maximize the coverage of stain styles. }
    \label{fig:method}
    %\vspace{-0.5cm}
\end{figure*}

\subsection{Overview}
Traditional classification approaches trained on single-scanner data demonstrate limited capability in handling diverse stain variations, leading to significant errors when encountering varying staining styles (Fig. 1(a)). While direct WSI-level augmentation might seem intuitive, the gigapixel nature of WSIs makes such approaches computationally intensive and storage-demanding, while still failing to capture the full spectrum of staining variations (Fig. 1(b)).

To address this, Latent Style Augmentation (LSA) framework is proposed to perform augmentation directly on WSI-level features (Fig. 1(c)). By eliminating redundant WSI generation and feature extraction, LSA enables efficient online augmentation. This approach allows comprehensive coverage of the stain space, significantly enhancing the model's robustness to diverse staining variations.

The core of LSA is the construction of a Stain Transformer that can perform style transformation on WSI features in the latent space. In the following sections, we discuss in detail the data preparation for training the Stain Transformer (Sec. \ref{sec:WSAug}), its architecture and training paradigm (Sec. \ref{sec:st}), as well as how it aids in the training of the WSI classifier (Sec. \ref{sec: LSA}).

\begin{figure*}[ht]
    \centering
    \includegraphics[width=\textwidth]{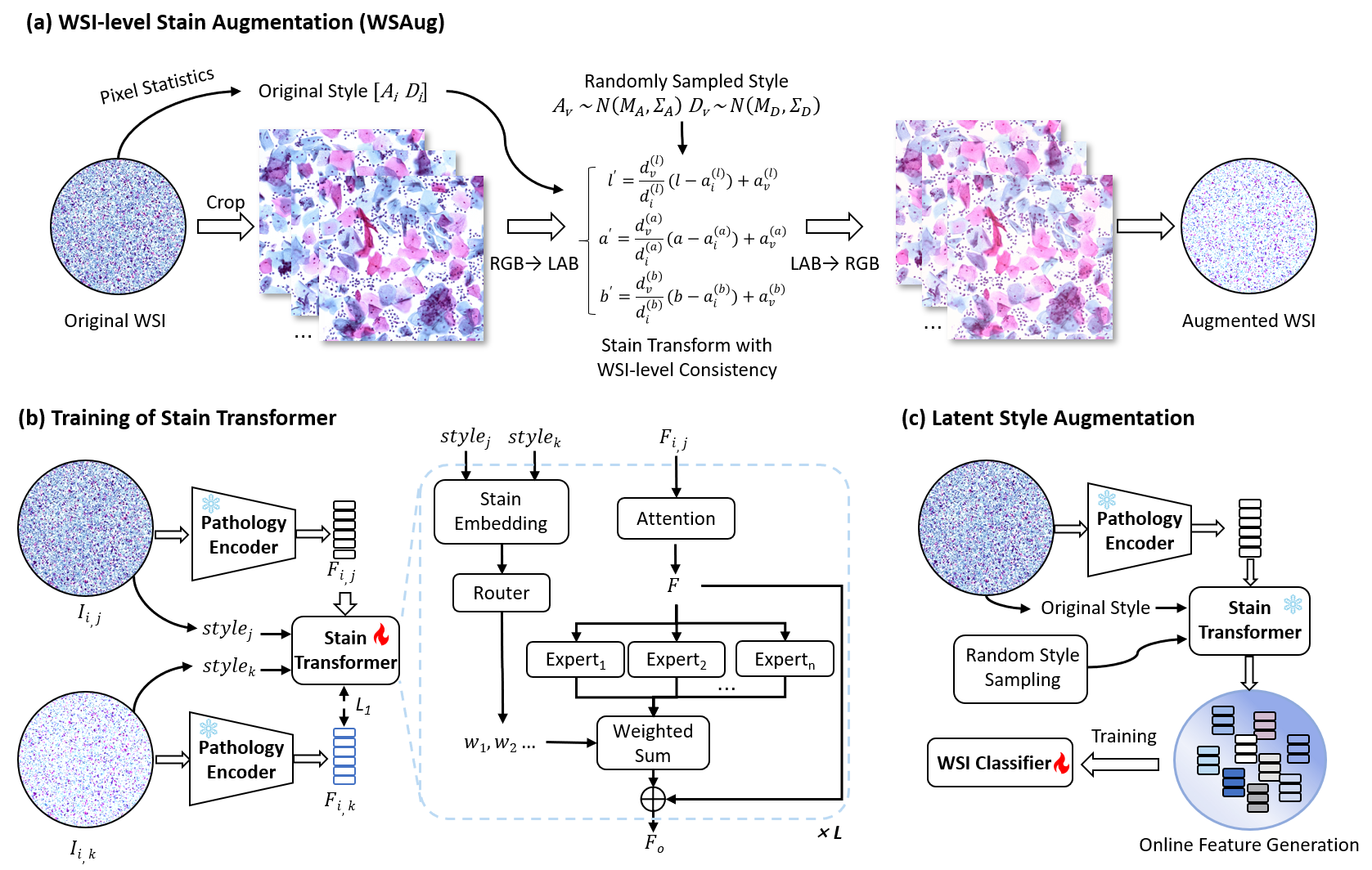}
    \caption{Implementation details. (a) The process of WSI-level stain augmentation (WSAug). (b) Training of Stain Transformer. (c) Latent Style Augmentation during training of MIL methods.}
    \label{fig:st}
    %\vspace{-0.5cm}
\end{figure*}

\subsection{WSI-level Stain Augmentation}
\label{sec:WSAug}
Since the Stain Transformer needs to learn the transformation across different styles, we first need to explicitly simulate high-quality stain transformations in varying styles.
Previous stain augmentation methods have been conducted at the patch level, with RandStainNA \cite{shen2022randstainna} being a particularly influential work in recent years.
However, directly applying patch-level stain augmentation to each patch of a WSI can lead to inconsistent styles within the WSI. 

To address this, we propose a stable WSI-level stain augmentation method, WSAug to optimize the consistency of WSI stain augmentation, as shown in Fig. \ref{fig:st}(a).
While the overall structure of WSAug remains in line with RandStainNA, the definitions of stain styles have been extended to the WSI-level.
Specifically, after cropping patches and converting the original WSI patches from RGB into LAB space, the pixel values are written as $[l, a, b]$.
We denote the average and standard deviation within channels for \textbf{all pixels of the WSI} as $\mathcal{A}_i=[a_i^{(l)},a_i^{(a)},a_i^{(b)}]$ and $\mathcal{D}_i=[d_i^{(l)},d_i^{(a)},d_i^{(b)}]$.
The styles of enhanced WSIs are denoted as $\mathcal{A}_v=[a_v^{(l)},a_v^{(a)},a_v^{(b)}]$ and $\mathcal{D}_v=[d_v^{(l)},d_v^{(a)},d_v^{(b)}]$, where $A_v$ and $D_v$ are sampled from Gaussian distribution built upon all the samples.
During a single stain transformation for a WSI, $[\mathcal{A}_i,\mathcal{D}_i]$ and $[\mathcal{A}_v,\mathcal{D}_v]$ are used for all cropped patches to generate $[l',a',b']$, and transfer pixels back to RGB space.
Taking the L channel as an example, the transformation process can be formulated as:

\begin{equation}
\begin{aligned}
l' &= \frac{d_v^{(l)}}{d_i^{(l)}}(l - a_i^{(l)}) + a_v^{(l)}.
\end{aligned}
\end{equation}

By applying three random WSAug transformations to each original WSI, we obtained WSIs in different styles to support the training of Stain Transformer.

% \begin{equation}
% \left\{
% \begin{aligned}
% l' &= \frac{d_v^{(l)}}{d_i^{(l)}}(l - a_i^{(l)}) + a_v^{(l)} \\
% a' &= \frac{d_v^{(a)}}{d_i^{(a)}}(a - a_i^{(a)}) + a_v^{(a)} \\
% b' &= \frac{d_v^{(b)}}{d_i^{(b)}}(b - a_i^{(b)}) + a_v^{(b)},
% \end{aligned}
% \right.
% \label{eq:WSAug}
% \end{equation}

\subsection{Architecture and Training of Stain Transformer}
\label{sec:st}

% 为了在latent space中得到多样性的特征，我们提出Stain Transformer，利用WSAug的来增强WSI-level features到
Compared with previous stain transform networks, Stain Transformer has two major advantages: it enables controllable transformations based on style input $s = [a^{(l)},a^{(a)},a^{(b)},d^{(l)},d^{(a)},d^{(b)}]$, rather than generating random or only target-domain data; and it directly operates on features, significantly reducing computational costs.
Details of Stain Transformer training are shown in Fig \ref{fig:st}(c). 

Before training, we extracted WSI-level features $F_{i,0...M}$ from both original WSI  $I_{i,0}$ and $M$ enhanced WSIs $I_{i,1...M}$. 
$F$ is directly formed by concatenating features of dimension $C$ extracted from $N$ cropped patches using the pre-trained encoder.
Stain Transformer generates corresponding features for enhanced WSIs, also maintaining the shape $[N,C]$.

Stain Transformer is composed of $L$ layers of transformer blocks, in which the Mixture of Experts (MoE) module \cite{jacobs1991adaptive} is controlled by the stain embedding. 
Stain Embedding block $SE(\cdot)$ embeds the concatenation of the original style $s_j$ and the randomly sampled style $s_k$ by a Linear layer. 
The MoE component consists of $n$ Feed-Forward Network (FFN) experts $E_i$. The route receives stain embedding as inputs, computing weights $w_{1...n}$ dynamically for each expert. 
The output from these experts is then aggregated through a weighted averaging process, effectively leveraging different experts’ configurations to cover wider scenarios. 
This design enhances Stain Transformer’s capacity to generate diverse and accurate features tailored to specific staining styles, formulated as:
\begin{equation}
\begin{aligned}
w_i &= \text{softmax}(\text{Linear}(SE(s_j,s_k)), \\
F_o &=  \text{GeLU}(\sum_{i=1}^{n} w_i \cdot E_i(F)),
\end{aligned}
\end{equation}
and the training of Stain Transformer $ST(\cdot)$ follows:
\begin{equation}
\begin{aligned}
L_{st} = L_1(ST(F_{i,j},s_j,s_k),F_{i,k}),
\end{aligned}
\end{equation}
where $j,k \in [0,M]$, because the source feature $F_{i,j}$ and target feature $F_{i,k}$ can be arbitrarily combined from the original and augmented data.

% After training is completed, Stain Transformer can be frozen and utilized for online random augmentation during the training of WSI classifier.

\subsection{ Latent Style Augmentation}
\label{sec: LSA}

LSA is employed to directly augment WSI-level latent features, as shown in Fig. \ref{fig:st}(c). In our framework, the model first extracts features from $N$ cropped patches of a WSI using a pre-trained pathology foundation model.
For each WSI, LSA calculates the original style and samples a target style in the same manner as WSAug in every step of training.
It then applies our designed Stain Transformer to perform online style transformation. The optimization of WSI-level classifier $f$ can be formulated as:

\begin{equation}
\begin{aligned}
L_{cls} = CE(f(ST(F_{i,j},s_j,s_k)),Y_i),
\end{aligned}
\end{equation}
where $CE$ denotes cross-entropy loss, $Y_i$ denotes the label of $I_i$.

% This approach enables the trained WSI-level classifier to maintain robustness across different styles of the same WSI, without the need to explicitly generate or repeatedly extract features from WSIs of the corresponding styles. 
% As a result, it significantly reduces computational and storage costs.

\section{Experiments}
\subsubsection{Datasets and Experimental Setup} 
We have collected a dataset consisting of 20,037 WSIs sourced from 18 different scanners, which are used for cervical cancer diagnosis. The hardware and parameter settings of different scanners introduce significant staining variations in the images they capture.
We selected the data from Scanner 1, which collects the most data, for training (2,353 positives, 1,804 negatives) and ID (in-domain) validation (565 positives, 691 negatives). 
The remaining data (6,419 positives, 8,205 negatives), from Scanner 2-18 are used for OOD (out-of-domain) validation.

To train Stain Transformer, we randomly applied WSAug 3 times to every WSI from Scanner 1, and pre-extract features for all WSIs.
The means and standard deviations of the Gaussian distributions are calculated from the datasets.
Besides, we obtained another set of augmented WSIs for the validation of Stain Transformer.
All experiments were optimized using Adam and conducted on an Nvidia A100-40G GPU. 
Stain Transformer is trained for 100 epochs with batch size of 8 and initial learning rate of $1\times 10^{-4}$, which is decreased by the StepLR scheduler with $\gamma=0.98$.
For all experiments of MIL methods, the WSI-level classifier is trained for 100 epochs with batch size of 32 and an initial learning rate of $1\times 10^{-4}$, which is decreased by the Cosine scheduler.

To demonstrate the generalization of our approach, we conduct validation using three influential pathology foundation models, including UNI \cite{chen2024towards}, Prov-GigaPath \cite{xu2024whole}, and H-optimus \cite{hoptimus0}, as patch-level encoders. 
Furthermore, we employed three popular MIL methods to conduct WSI-level classification, including TransMIL \cite{shao2021transmil}, ABMIL \cite{ilse2018attention}, and MambaMIL \cite{yang2024mambamil}. 
We conducted systematic experiments with various model integrations, thereby verifying the robustness.

\begin{table}[b]
\caption{Comparison between different methods based on different foundational models. In this table, WSAug and SCLA are built upon TransMIL.}
\centering
\begin{tabular}{c|cccccc|cccccc}
\hline
Scanner & \multicolumn{6}{c|}{Scanner 1 (ID)} & \multicolumn{6}{c}{Scanner 2-18 (OOD)} \\ \hline
Encoder & \multicolumn{2}{c}{UNI} & \multicolumn{2}{c}{GigaPath} & \multicolumn{2}{c|}{H-optimus} & \multicolumn{2}{c}{UNI} & \multicolumn{2}{c}{GigaPath} & \multicolumn{2}{c}{H-optimus} \\ 
Metrics& AUC & ACC & AUC & ACC& AUC & ACC& AUC & ACC& AUC & ACC& AUC & ACC \\
\hline
TransMIL & 85.53 & 80.42 & 84.52 & 77.33 & 87.79 & 81.85 & 77.52 & 68.87 & 76.78 & 69.19 & 82.22 & 64.12 \\
ABMIL & 85.37 & 79.54 & 86.75 & 79.06 & 88.75 & 82.16 & 77.00 & 65.53 & 79.88 & 67.28 & 82.10 & 62.11 \\
MambaMIL & 86.01 & 80.01 & 86.41 & 80.10 & 87.81 & 81.68 & 78.26 & 71.47 & 78.17 & 68.12 & 83.10 & 65.45 \\
\pmb{WSAug (Ours)} & 85.61 & 80.10 & \pmb{87.44} & 80.49 & \pmb{88.86} & \pmb{83.20} & 81.97 & 70.06 & 83.82 & 70.45 & 84.56 & 71.51 \\
\pmb{SCLA (Ours)} & \pmb{87.09} & \pmb{80.81} & 86.82 & \pmb{80.97} & 88.41 & 81.29 & \pmb{85.85} & \pmb{72.66} & \pmb{84.71} & \pmb{78.26} & \pmb{86.11} & \pmb{73.52} \\ \hline

\end{tabular}
\label{tab:compare}
\end{table}

\subsubsection{Comparison with Other MIL Methods} 
Table \ref{tab:compare} presents the original performance of each comparative method as well as the optimized performance achieved by combining TransMIL with our method. 
We can draw significant observations from these results:
(1) The performance of each comparative method on the OOD scanners is far less effective than on the ID scanner. 
(2) The performance of merely introducing augmented data from WSAug remains stable on the ID dataset, but there is a remarkable improvement on the OOD dataset. 
(3) LSA further boosts performance on OOD validation, outperforming other methods across all combinations of various encoders and MIL methods.

\begin{table}[ht]

\caption{We compare our methods with other augmentation methods (SN \cite{tellez2019quantifying}, SA \cite{946629}, RandStainNA \cite{shen2022randstainna}). All results are based on features extracted by GigaPath.}
\centering
\begin{tabular}{c|cccccc|cccccc}
\hline
Scanner & \multicolumn{6}{c|}{Scanner 1 (ID)} & \multicolumn{6}{c}{Scanner 2-18 (OOD)} \\ \hline
Method & \multicolumn{2}{c}{TransMIL} & \multicolumn{2}{c}{ABMIL} & \multicolumn{2}{c|}{MambaMIL} & \multicolumn{2}{c}{TransMIL} & \multicolumn{2}{c}{ABMIL} & \multicolumn{2}{c}{MambaMIL} \\ 
Metrics & AUC & ACC & AUC & ACC & AUC & ACC & AUC & ACC & AUC & ACC & AUC & ACC \\ 
\hline
Baseline & 84.52 & 77.33 & 86.75 & 79.06 & 86.41 & 80.10 & 76.78 & 69.19 & 79.88 & 67.28 & 78.17 & 68.12 \\
SN & 85.04 & 77.78 & 86.81 & 79.45 & 86.44 & 80.73 & 65.54 & 59.64 & 71.87 & 59.09 & 70.62 & 59.52 \\
SA & 84.95 & 77.85 & 86.32 & 79.77 & 84.41 & 80.17 & 80.63 & 69.21 & 75.97 & 61.90 & 78.71 & 68.54\\
RandStainNA & 85.00 & 77.70 & 85.39 & 79.86 & 86.80 & 79.45 & 78.52 & 69.16 & 77.45 & 60.23 & 79.60 & 69.50 \\
\pmb{WSAug (Ours)} & \pmb{87.44} & 80.49 & 87.08 & 80.57 & \pmb{87.36} & 80.97 &  83.82& 70.45 & 82.95 & 65.73 & 83.26 & 71.70 \\
\pmb{SCLA (Ours)} & 86.82 & \pmb{80.97} & \pmb{87.67} & \pmb{81.61} & 87.31 & \pmb{82.40} & \pmb{84.71} & \pmb{78.26} & \pmb{84.10} & \pmb{69.42} & \pmb{84.18} & \pmb{73.80} \\
\hline
\end{tabular}
\label{tab:ablation}
\end{table}

\subsubsection{Comparison with Other Augmentation Methods} 
Table \ref{tab:ablation} compares our method with different patch-level stain augmentation methods, including Stain Normalization (SN) \cite{tellez2019quantifying}, Stain Augmentation (SA) \cite{946629}, and RandStainNA \cite{shen2022randstainna}.
To the best of our knowledge, there are currently no direct WSI-level stain augmentation methods available for comparison. 
The results are obtained by using UNI as the encoder and various MIL methods as the WSI-level aggregator. 
We have observed that directly applying patch-level augmentation methods to WSI classification does not enhance cross-scanner performance and may even exacerbate overfitting on the training scanner in some cases. 
We attribute this to the inability of simple patch-level processing to maintain consistent stain style across all patches within a WSI. 
This inconsistency undermines the generalizability and robustness of the WSI classifier to different scanners.

\begin{table}[b]

\caption{Classification performance after adding augmented data of varying sizes using WSAug based on TransMIL. }
\centering
\begin{tabular}{c|cccccc|cccccc}
\hline
Scanner & \multicolumn{6}{c|}{Scanner 1 (ID)} & \multicolumn{6}{c}{Scanner 2-18 (OOD)} \\ \hline
Encoder & \multicolumn{2}{c}{UNI} & \multicolumn{2}{c}{GigaPath} & \multicolumn{2}{c|}{H-optimus} & \multicolumn{2}{c}{UNI} & \multicolumn{2}{c}{GigaPath} & \multicolumn{2}{c}{H-optimus} \\ 
Metrics & AUC & ACC & AUC & ACC & AUC & ACC & AUC & ACC & AUC & ACC & AUC & ACC \\ 
\hline
Baseline & 85.53 & 80.42 & 84.52 & 77.33 & 87.79 & 81.85 & 77.52 & 68.87 & 76.78 & 69.19 & 82.22 & 64.12 \\
+100$\%$ & 85.36 & 80.35 & 84.91 & 78.92 & 87.69 & 81.77 & 81.00 & 68.54 & 80.34 & 69.63 & 83.78 & 65.93 \\
+200$\%$ & \pmb{86.54} & \pmb{80.83} & 84.29 & 77.95 & 85.70 & 80.02 & 81.35 & 69.85 & 81.31 & 70.08  & 84.83 & 70.25  \\
+300$\%$ & 85.61 & 80.10 & \pmb{87.44} & \pmb{80.49} & \pmb{88.86} & \pmb{83.20} & \pmb{81.97} & \pmb{70.06} & \pmb{83.82} & \pmb{70.45} & \pmb{84.56}  & \pmb{71.51}  \\

\hline
\end{tabular}
\label{tab:WSAug}
\end{table}

\subsubsection{Influence of Augmented Data Size} Table \ref{tab:WSAug} presents the performance of various MIL methods after adding augmented data with varying sizes using WSAug. 
The models exhibit somewhat unstable classification results on the ID dataset when increasing augmented data size, yet they generally perform slightly better than the baseline. 
In contrast, on the OOD dataset, the model performance shows a noticeable improvement with the expansion of augmented data size.
Although the performance improvement brought by data augmentation has not yet reached its upper limit, the substantial time costs have temporarily prevented us from further exploration. Augmenting all WSIs in the training set takes 80, 120, and 120 hours for UNI, GigaPath, and H-optimus, respectively.

\subsubsection{Effectiveness of Stain Transformer} 
% We employed the Mean Absolute Error (MAE) between feature matrices as a metric to evaluate the effectiveness of the Stain Transformer. 
% The MAE values, which represent the discrepancies between the features extracted by the base models (UNI, GigaPath, and H-optimus) for the input and target, were initially recorded as follows: 1.01 for UNI, 0.7109 for GigaPath, and 0.5921 for H-optimus. 
The Mean Absolute Error (MAE) between features can represent the discrepancies between WSIs with different staining styles.
To evaluate the performance of Stain Transformer, we calculated MAE between the features of the offline augmented WSIs and the prediction of Stain Transformer.
Before the process of Stain Transformer, the MAE values are 1.01 for UNI, 0.7109 for GigaPath, and 0.5921 for H-optimus.
After applying the Stain Transformer, the MAE values were significantly reduced to 0.4478 for UNI, 0.3142 for GigaPath, and 0.2619 for H-optimus.
The results indicate that Stain Transformer is capable of generating WSI-level features corresponding to the target stain style, enabling online stain augmentation for WSI-level features.

\subsubsection{Influence of Augmentation Possibility} 
Table \ref{tab:generate} illustrates the impact of the possibility $p$ of applying LSA during training. 
It can be observed that across various settings, LSA consistently enhances the classification performance of different models compared to no augmentation ($p=0$). 
When $p$ is set to 0.5, the models achieve the best balance in performance on both ID and OOD datasets.

\begin{table}[h]

\caption{Classification performance after Stain-Conditioned Feature Augmentation based on TransMIL. $p$ represents the possibility of augmentation during training. }
\centering
\begin{tabular}{c|cccccc|cccccc}
\hline
Scanner & \multicolumn{6}{c|}{Scanner 1 (ID)} & \multicolumn{6}{c}{Scanner 2-18 (OOD)} \\ \hline
Encoder & \multicolumn{2}{c}{UNI} & \multicolumn{2}{c}{GigaPath} & \multicolumn{2}{c|}{H-optimus} & \multicolumn{2}{c}{UNI} & \multicolumn{2}{c}{GigaPath} & \multicolumn{2}{c}{H-optimus} \\ 
Metrics & AUC & ACC & AUC & ACC & AUC & ACC & AUC & ACC & AUC & ACC & AUC & ACC \\ 
\hline
$p=0$ & 85.53 & 80.42 & 84.52 & 77.33 & 87.79 & 81.85 & 77.52 & 68.87 & 76.78 & 69.19 & 82.22 & 64.12 \\
$p=0.2$  & \pmb{88.11} & \pmb{82.72} & 85.73 & 78.82 & 88.41 & 81.29 & 85.44 & 69.58 & 83.56 & 73.12 & 85.46 & 72.56 \\
$p=0.5$ & 87.09 & 80.81 & \pmb{86.82} & \pmb{80.87} & \pmb{88.88} & \pmb{82.24} & \pmb{85.85} & \pmb{72.66} & \pmb{84.71} & \pmb{78.26} & \pmb{86.11} & \pmb{73.52} \\
$p=0.8$  & 85.41 & 77.87 & 84.43 & 78.03 & 88.18 & 80.97 & 85.15 & 67.16 & 82.12 & 69.88 & 85.97 & 73.06 \\

\hline
\end{tabular}
\label{tab:generate}
\end{table}

\section{Conclusion}

% In conclusion, this study extends the widely utilized patch-level staining augmentation method RandStainNA to the Whole Slide Image (WSI) level. We introduce a novel staining augmentation method, style-conditioned latent augmentation, which is capable of augmenting classifier training online, effectively transforming WSI-level features to correspond with various staining styles. Our collected cervical cancer WSI diagnosis dataset, which includes data from multiple scanners, demonstrates a significant performance improvement when trained on data from one scanner and validated on others. Future work will focus on conducting experiments on publicly available histopathology datasets to validate our approach further.

In this work, we address the challenge of domain shifts in WSI-based cervical cancer diagnosis by designing a novel stain augmentation paradigm. 
By operating directly in the WSI-level feature space, the proposed LSA bypasses the computational bottlenecks of traditional stain augmentation methods, achieving online augmentation at the WSI level.
Experiments on a multi-scanner dataset demonstrate the effectiveness of our approach.

While our method demonstrates promising results, we identify the key limitations that warrant further investigation. 
The training of Stain Transformer currently depends on limited offline generated WSIs. 
Although these synthetic variations improve cross-scanner generalization, they may not fully capture the real-world interactions between staining protocols and scanner hardware.
Addressing this gap will require integrating dynamic, physics-based stain simulation frameworks into the training process. 

% Nonetheless, we acknowledge the limitations of the current approach. The training of the Stain Transformer still relies on pre-simulated stain transformations performed offline, which means there is still some distance to go before we can fully and accurately simulate stain transformations, necessitating further technological exploration. Additionally, our method has thus far only been validated on cervical cancer data, and we hope to further validate it in histopathology in the future.
%
%


\bibliographystyle{unsrt}
\bibliography{main}

\begin{thebibliography}{10}

\bibitem{gultekin2020world}
MURAT G{\"U}LTEK{\.I}N, Pedro Ramirez, Nathalie Broutet, and Raymond Hutubessy.
\newblock World health organization call for action to eliminate cervical cancer globally, 2020.

\bibitem{cheng2021robust}
Shenghua Cheng, Sibo Liu, Jingya Yu, Gong Rao, Yuwei Xiao, Wei Han, Wenjie Zhu, Xiaohua Lv, Ning Li, Jing Cai, et~al.
\newblock Robust whole slide image analysis for cervical cancer screening using deep learning.
\newblock {\em Nature communications}, 12(1):5639, 2021.

\bibitem{wang2024artificial}
Jue Wang, Yunfang Yu, Yujie Tan, Huan Wan, Nafen Zheng, Zifan He, Luhui Mao, Wei Ren, Kai Chen, Zhen Lin, et~al.
\newblock Artificial intelligence enables precision diagnosis of cervical cytology grades and cervical cancer.
\newblock {\em Nature Communications}, 15(1):4369, 2024.

\bibitem{jiang2023donet}
Hao Jiang, Rushan Zhang, Yanning Zhou, Yumeng Wang, and Hao Chen.
\newblock Donet: Deep de-overlapping network for cytology instance segmentation.
\newblock In {\em Proceedings of the IEEE/CVF conference on computer vision and pattern recognition}, pages 15641--15650, 2023.

\bibitem{cai2023progressive}
Jiangdong Cai, Honglin Xiong, Maosong Cao, Luyan Liu, Lichi Zhang, and Qian Wang.
\newblock Progressive attention guidance for whole slide vulvovaginal candidiasis screening.
\newblock In {\em International Conference on Medical Image Computing and Computer-Assisted Intervention}, pages 233--242. Springer, 2023.

\bibitem{shao2021transmil}
Zhuchen Shao, Hao Bian, Yang Chen, Yifeng Wang, Jian Zhang, Xiangyang Ji, et~al.
\newblock Transmil: Transformer based correlated multiple instance learning for whole slide image classification.
\newblock {\em Advances in neural information processing systems}, 34:2136--2147, 2021.

\bibitem{yang2024mambamil}
Shu Yang, Yihui Wang, and Hao Chen.
\newblock Mambamil: Enhancing long sequence modeling with sequence reordering in computational pathology.
\newblock In {\em International Conference on Medical Image Computing and Computer-Assisted Intervention}, pages 296--306. Springer, 2024.

\bibitem{ilse2018attention}
Maximilian Ilse, Jakub Tomczak, and Max Welling.
\newblock Attention-based deep multiple instance learning.
\newblock In {\em International conference on machine learning}, pages 2127--2136. PMLR, 2018.

\bibitem{hoptimus0}
Charlie Saillard, Rodolphe Jenatton, Felipe Llinares-López, Zelda Mariet, David Cahané, Eric Durand, and Jean-Philippe Vert.
\newblock H-optimus-0, 2024.

\bibitem{huang2023visual}
Zhi Huang, Federico Bianchi, Mert Yuksekgonul, Thomas~J Montine, and James Zou.
\newblock A visual--language foundation model for pathology image analysis using medical twitter.
\newblock {\em Nature medicine}, 29(9):2307--2316, 2023.

\bibitem{chen2024towards}
Richard~J Chen, Tong Ding, Ming~Y Lu, Drew~FK Williamson, Guillaume Jaume, Andrew~H Song, Bowen Chen, Andrew Zhang, Daniel Shao, Muhammad Shaban, et~al.
\newblock Towards a general-purpose foundation model for computational pathology.
\newblock {\em Nature Medicine}, 30(3):850--862, 2024.

\bibitem{li2024large}
Honglin Li, Yusuan Sun, Chenglu Zhu, Yunlong Zhang, Shichuan Zhang, Zhongyi Shui, Pingyi Chen, Jingxiong Li, Sunyi Zheng, Can Cui, et~al.
\newblock Large-scale cervical precancerous screening via ai-assisted cytology whole slide image analysis.
\newblock {\em arXiv preprint arXiv:2407.19512}, 2024.

\bibitem{cao2023detection}
Maosong Cao, Manman Fei, Jiangdong Cai, Luyan Liu, Lichi Zhang, and Qian Wang.
\newblock Detection-free pipeline for cervical cancer screening of whole slide images.
\newblock In {\em International Conference on Medical Image Computing and Computer-Assisted Intervention}, pages 243--252. Springer, 2023.

\bibitem{jahanifar2023domaingeneralizationcomputationalpathology}
Mostafa Jahanifar, Manahil Raza, Kesi Xu, Trinh Vuong, Rob Jewsbury, Adam Shephard, Neda Zamanitajeddin, Jin~Tae Kwak, Shan E~Ahmed Raza, Fayyaz Minhas, and Nasir Rajpoot.
\newblock Domain generalization in computational pathology: Survey and guidelines, 2023.

\bibitem{salehi2020pix2pix}
Pegah Salehi and Abdolah Chalechale.
\newblock Pix2pix-based stain-to-stain translation: A solution for robust stain normalization in histopathology images analysis.
\newblock In {\em 2020 International Conference on Machine Vision and Image Processing (MVIP)}, pages 1--7. IEEE, 2020.

\bibitem{bentaieb2017adversarial}
A{\"\i}cha BenTaieb and Ghassan Hamarneh.
\newblock Adversarial stain transfer for histopathology image analysis.
\newblock {\em IEEE transactions on medical imaging}, 37(3):792--802, 2017.

\bibitem{shaban2019staingan}
M~Tarek Shaban, Christoph Baur, Nassir Navab, and Shadi Albarqouni.
\newblock Staingan: Stain style transfer for digital histological images.
\newblock In {\em 2019 Ieee 16th international symposium on biomedical imaging (Isbi 2019)}, pages 953--956. IEEE, 2019.

\bibitem{wagner2021structure}
Sophia~J Wagner, Nadieh Khalili, Raghav Sharma, Melanie Boxberg, Carsten Marr, Walter De~Back, and Tingying Peng.
\newblock Structure-preserving multi-domain stain color augmentation using style-transfer with disentangled representations.
\newblock In {\em Medical Image Computing and Computer Assisted Intervention--MICCAI 2021: 24th International Conference, Strasbourg, France, September 27--October 1, 2021, Proceedings, Part VIII 24}, pages 257--266. Springer, 2021.

\bibitem{shen2022randstainna}
Yiqing Shen, Yulin Luo, Dinggang Shen, and Jing Ke.
\newblock Randstainna: Learning stain-agnostic features from histology slides by bridging stain augmentation and normalization.
\newblock In {\em International Conference on Medical Image Computing and Computer-Assisted Intervention}, pages 212--221. Springer, 2022.

\bibitem{zhang2023learning}
Zuyu Zhang, Yan Li, and Byeong-Seok Shin.
\newblock Learning generalizable visual representation via adaptive spectral random convolution for medical image segmentation.
\newblock {\em Computers in Biology and Medicine}, 167:107580, 2023.

\bibitem{jacobs1991adaptive}
Robert~A Jacobs, Michael~I Jordan, Steven~J Nowlan, and Geoffrey~E Hinton.
\newblock Adaptive mixtures of local experts.
\newblock {\em Neural computation}, 3(1):79--87, 1991.

\bibitem{xu2024whole}
Hanwen Xu, Naoto Usuyama, Jaspreet Bagga, Sheng Zhang, Rajesh Rao, Tristan Naumann, Cliff Wong, Zelalem Gero, Javier Gonz{\'a}lez, Yu~Gu, et~al.
\newblock A whole-slide foundation model for digital pathology from real-world data.
\newblock {\em Nature}, pages 1--8, 2024.

\bibitem{tellez2019quantifying}
David Tellez, Geert Litjens, P{\'e}ter B{\'a}ndi, Wouter Bulten, John-Melle Bokhorst, Francesco Ciompi, and Jeroen Van Der~Laak.
\newblock Quantifying the effects of data augmentation and stain color normalization in convolutional neural networks for computational pathology.
\newblock {\em Medical image analysis}, 58:101544, 2019.

\bibitem{946629}
E.~Reinhard, M.~Adhikhmin, B.~Gooch, and P.~Shirley.
\newblock Color transfer between images.
\newblock {\em IEEE Computer Graphics and Applications}, 21(5):34--41, 2001.

\end{thebibliography}





% LNCS Homepage, \url{http://www.springer.com/lncs}, last accessed 2023/10/25

\end{document}